\documentclass[aps,prl,twocolumn,preprintnumbers,amsmath,amssymb,superscriptaddress]{revtex4-2}

\usepackage{amssymb}
\usepackage{mathtools}
\usepackage{physics}
\usepackage{graphicx}
\usepackage{hyperref}
\usepackage{slashed}
\usepackage{orcidlink}
\usepackage{bbm}
\usepackage{mathtools}

\DeclarePairedDelimiter\floor{\lfloor}{\rfloor}
\usepackage{color}
\usepackage[normalem]{ulem}
\usepackage[export]{adjustbox}

\begin{document}

\title{UV complete local field theory of persistent symmetry breaking in 2+1 dimensions}

\author{Bilal Hawashin}
\email{hawashin@tp3.rub.de}
\affiliation{Institut f\"ur Theoretische Physik III, Ruhr-Universit\"at Bochum, D-44801 Bochum, Germany}

\author{Junchen Rong}
\email{junchenrong@ihes.fr}
\affiliation{Institut des Hautes \'Etudes Scientifiques, 91440 Bures-sur-Yvette, France}

\author{Michael M. Scherer}
\email{scherer@tp3.rub.de}
\affiliation{Institut f\"ur Theoretische Physik III, Ruhr-Universit\"at Bochum, D-44801 Bochum, Germany}

\begin{abstract}
Spontaneous symmetry breaking can persist at all temperatures in certain biconical $\mathrm{O}(N)\times \mathbb{Z}_2$ vector models when the underlying field theories are ultraviolet complete.
So far, the existence of such theories has been established in fractional dimensions for local but nonunitary models or in 2+1 dimensions but for nonlocal models.
Here, we study local models at zero and finite temperature directly in 2+1 dimensions employing functional methods.
At zero temperature, we establish that our approach describes the quantum critical behaviour with good accuracy for all $N\geq 2$.
We then exhibit the mechanism of discrete symmetry breaking from $\mathrm{O}(N)\times \mathbb{Z}_2\to \mathrm{O}(N)$ for increasing temperature near the biconical critical point when $N$ is finite but large.
We calculate the corresponding finite-temperature phase diagram and further show that the Hohenberg-Mermin-Wagner theorem is fully respected within this approach, i.e., symmetry breaking only occurs in the $\mathbb{Z}_2$ sector.
Finally, we determine the critical $N$ above which this phenomenon can be observed to be $N_c \approx 15$.
\end{abstract}

\maketitle


\textit{Introduction. ---} 
At high temperatures, any system in thermodynamic equilibrium has to be in a phase with high entropy $S$ to minimize the free energy ${F = E - TS}$. 
Typically, states with high entropy are disordered and, if spontaneous symmetry-breaking (SSB) is involved, this usually means that the disordered high-temperature phase has higher symmetry.
An exception is the Pomeranchuk effect~\cite{pomeranchuk1950theory} in Helium-3 at very low temperatures, which occurs in a narrow region of the phase diagram: Here, liquid Helium-3 crystallizes when heated, owing to an excess entropy of spins in the solid phase~\cite{RevModPhys.69.645}.
Upon further heating, the Helium melts, again, restoring the symmetry of the liquid phase. 
Related effects can also be observed in moir\'e materials~\cite{rozen2021entropic,saito2021isospin,li2021continuous} and Rochelle salt~\cite{4055358}.
Such `inverted' phase diagrams, where SSB occurs towards higher temperatures, can also emerge in quantum field theoretical (QFT) models 
with extra degrees of freedom accompanying the formation of order~\cite{Weinberg:1974hy,BIMONTE1996248,PIETRONI1997119,ORLOFF1997309,PhysRevD.54.2944,PhysRevD.61.125016,PhysRevD.103.096014}.
The non-restoration of their symmetry at high temperatures has inspired applications in cosmology~\cite{Mohapatra:1979qt,PhysRevD.20.3390,PhysRevLett.45.1,PhysRevD.21.3470,Salomonson:1984rh,PhysRevLett.64.340,KUZMIN1981159,KUZMIN1981167,Dodelson:1991iv,Dvali:1995cj,PhysRevD.54.7153,PhysRevLett.122.041802}. 
An obvious question is, if such phenomena can persist at arbitrarily high temperature -- in contrast to common expectations. 
Due to the lack of ultraviolet~(UV) completeness of many QFTs, however, there is typically an upper cutoff, limiting the range of viable temperatures.

Recently, UV complete QFT models with SSB at all temperatures $T$ have been suggested in  $D=4-\epsilon$~\cite{PhysRevLett.125.131603, Chai:2020zgq}.
The work was followed by studies of related long-range models in $D\!=\!2\!+\!1$~\cite{Chai:2021djc}, by four-dimensional scenarios, i.e., large $N$ gauge theory~\cite{Chaudhuri:2020xxb,Chaudhuri:2021dsq} and asymptotic safety~\cite{PhysRevD.103.096014}.
Non-unitary models with random chemical potential had been discussed, previously~\cite{PhysRevD.63.085014}.
In~\cite{PhysRevLett.125.131603, Chai:2020zgq}, the UV completion corresponds to the existence of a quantum critical point~(QCP) defining a conformal field theory~(CFT).
The inverted phase transition then occurs near the QCP, given that the corresponding CFT at finite $T$ lies in the SSB phase.
Notably, a CFT at finite temperature has no intrinsic scale except $T$ and a dimensional analysis reveals that the $T$~dependence of the vacuum expectation value of an order parameter $\phi$ with scaling dimension $\Delta_{\phi}$ is $\langle\phi\rangle = b_{\phi} T^{\Delta_{\phi}}$~\cite{Katz:2014rla, Iliesiu:2018fao}. Here $b_\phi$ is a non-universal dimensionless constant. When $b_{\phi} \neq 0$, the system is in the SSB phase for $T>0$. This relation is valid up to arbitrarily high temperatures~\footnote{The dynamical mechanism of persistent order discussed here should be distinguished from another mechanism due to t'Hooft anomalies involving 1-form symmetries \cite{Komargodski:2017dmc,Tanizaki:2017qhf,Ciccone:2023pdk}. The latter mechanism does not lead to inverted phase diagrams, since the anomaly does not allow the trivially gapped disordered phase.}.  Hence, if a CFT shows SSB at some fixed finite temperature, then SSB can persist at all temperatures~\cite{Chai:2020zgq}, 
resulting in the phenomenon of ``persistent symmetry breaking''.

While Refs.~\cite{PhysRevLett.125.131603, Chai:2020zgq} provided an intriguing new direction for such persistent SSB, some key issues were left open.
Specifically, $\mathrm{O}(N)\times\mathrm{O}(M)$ symmetric models of two coupled scalars were studied, suggesting the existence of CFTs where one of the orthogonal groups is symmetry broken.
However, non-integer dimension $D=4-\epsilon$ leads to unitarity violation~\cite{PhysRevD.93.125025}.
Another issue appears when extrapolating to $\epsilon\!=\!1$, which corresponds to physics in two spatial dimensions at $T=0$.
At finite~$T$, the Coleman-Hohenberg-Mermin-Wagner theorem~(CHMW)~\cite{PhysRevLett.17.1133,PhysRev.158.383,Coleman:1973ci} forbids SSB of continuous symmetry $\mathrm{O}(M)$, leaving the exception of $M=1$, as $\mathrm{O}(1)\cong\mathbb{Z}_2$ is discrete. 
The above subtlety is non-perturbative and does not manifest itself in $\epsilon$~expansion, so the validity of the extrapolation to $\epsilon=1$ even for $M=1$ is questionable. 
A non-local version of the above models formulated directly in $2+1$ dimensions has been shown to feature persistent SSB by using long-range perturbation theory~\cite{Chai:2021djc,Chai:2021tpt}. 
Despite being UV complete, continuation to local models is not \textit{a priori} well-defined, as non-local theories are excluded from CHMW.

Here, we address these open issues and add substantial evidence that unitary and local $\mathrm{O}(N) \times \mathbb{Z}_2$ models at $N>N_c$ possess a UV completion facilitating persistent SSB. 
To this end, we employ the non-perturbative functional renormalization group (FRG) directly in $2+1$ dimensions. 
We explicitly calculate finite-$T$ phase diagrams near the biconical fixed point realizing a QCP, transparently expose the mechanism of inverted SSB in the functional approach, and show how CHMW is faithfully respected at the same time. We further provide an estimate of $N_c$ above which persistent SSB exists.


\textit{Model and method. ---} We consider a QFT in $D=d+1$-dimensional Euclidean spacetime with global symmetry $\mathrm{O}(N) \times \mathbb{Z}_2$, where $\phi=(\phi_1,\ldots,\phi_N)$ transforms as a vector under $\mathrm{O}(N)$ and $\chi$ changes sign under $\mathbb{Z}_2$. Invariants under the global symmetry are combinations of $\rho_\phi = \frac{1}{2} \phi_a \phi_a$ and $\rho_\chi = \frac{1}{2} \chi^2$.
The minimal action reads
\begin{align}\label{action}
    S =\!\int\! d^Dx \Big( \frac{1}{2}(\partial \phi)^2 + \frac{1}{2} (\partial \chi)^2 + U[\rho_\phi,\rho_\chi] \Big)\,,
\end{align}
with the scalar potential
\begin{align}\label{eq:micpot}
 U= m_\phi^2 \rho_\phi + m_\chi^2 \rho_\chi +\frac{\lambda_\phi}{2} \rho_\phi^2 + \frac{\lambda_\chi}{2} \rho_\chi^2 +\lambda_{\phi \chi} \rho_\phi \rho_\chi \,,   
\end{align}
which is bounded from below for $\lambda_{\phi},\lambda_{\chi} \geq 0$ and ${\lambda_\phi \lambda_\chi \geq \lambda_{\phi \chi}^2}$, allowing negative $\lambda_{\phi \chi}$.


A $D$-dimensional QFT at $T=0$ is promoted to a thermal field theory by compactifying the time domain on a circle with circumference $\beta = 1/T$, yielding a $T>0$ theory in $d$ spatial dimensions. 
For scalar QFTs, this is realized in momentum space by replacing
\begin{equation}
    q_0 \to i\omega_n,\quad \int \frac{d^D q}{(2\pi)^D} \to T \sum_{n \in \mathbb{Z}} \int \frac{d^{d} q}{(2\pi)^{d}},
\end{equation}
with bosonic Matsubara frequencies $\omega_n=2 \pi n T$.

To investigate the UV and thermodynamic behaviour of Eq.~\eqref{action}, we utilize FRG which facilitates, e.g., the study of universal critical phenomena, non-universal phase diagrams, finite temperature, symmetry-broken phases, and other
non-perturbative effects, see~\cite{Dupuis:2021} for a review. 
In FRG, the RG scale $k$ is introduced as an infrared~(IR) cut-off, modifying the path integral as
\begin{equation}\label{eq:modpath}
    \int[d\varphi] e^{-S[\varphi]} \to \int[d\varphi] e^{-S[\varphi] - \Delta S_k}\,.
\end{equation}
Here, $\Delta S_k$ is a functional of fields $\varphi$ and depends on a cut-off scheme $R_k$ acting as IR regulator.
Up to a small set of constraints, it can be chosen freely~\cite{Pawlowski:2005xe,Dupuis:2021, Hawashin:2023nordic}.
Eq.~\eqref{eq:modpath} then defines an effective action, modified by the regulator insertion, known as the flowing action~$\Gamma_k[\Phi]$ with $\Phi$ being the expectation value of $\varphi$ in the presence of sources~\cite{Dupuis:2021}. Removal of the cut-off yields the full effective action, i.e, $\lim_{k\to0} \Gamma_k = \Gamma$, which is the standard tool to study SSB~\cite{PhysRev.127.965,Jona-Lasinio:1964zvf,PhysRevD.7.1888}.
For a regulator insertion bilinear in~$\varphi$, i.e., 
$\Delta S_k[\varphi,R] = \int \frac{d^Dp}{(2\pi)^D} \varphi_\alpha(p) R_k^{\alpha \beta}(p^2) \varphi_\beta(-p)$,
the flow of $\Gamma_k$ is given by the Wetterich equation~\cite{Wetterich:1992yh},
\begin{equation}\label{eq:floweq}
    \partial_t \Gamma_k = \frac{1}{2} \mathrm{Tr} \left[ (\Gamma^{(2)}_k + R_k)^{-1} \partial_t R_k \right],
\end{equation}
where the trace runs over all field-indices, including spacetime coordinates or momenta/frequencies. Further, ${t = \log k /\Lambda}$ with UV cut-off $\Lambda$ at which the bare action is defined, and ${(\Gamma^{(2)}_k)^{\alpha \beta} = \frac{\delta^2}{\delta \Phi_\alpha \delta \Phi_\beta} \Gamma_k}$.
Eq.~\eqref{eq:floweq} is a regularized and non-perturbative one-loop equation. The flowing action $\Gamma_k$ can then be expanded in a series of local operators, which has to be truncated in practical calculations.


\textit{Renormalization group equations. ---} A suitable truncation scheme for our purposes is the extended local potential approximation~(LPA${}'$)
\begin{equation}
    \Gamma_k = \int d^Dx \Big[ \frac{Z_\phi}{2}(\partial \phi)^2 + \frac{Z_\chi}{2} (\partial \chi)^2 + U_k[\rho_\phi,\rho_\chi] \Big],
\end{equation}
where we include uniform field renormalizations~$Z_{\phi,\chi}$.
Related models have been studied successfully with FRG, see, e.g., Refs.~\cite{Bornholdt:1995,Bornholdt:1996,PIETRONI1997119,Eichhorn:2013mcl,PhysRevE.91.062112,khan2015phase,Borchardt:2016,PhysRevB.93.125119,Torres:2020}.
Employing the linear regulator $R_k^{\phi,\chi}(p^2) = Z_{\phi,\chi}(k^2 - p^2)\Theta(k^2-p^2)$, the flow of the dimensionless potential $u=k^{-D} U_k$ in terms of the invariants $\bar{\rho}_i = Z_ik^{2-D}\rho_i, i\in\{\phi,\chi\}$ is derived from Eq.~\eqref{eq:floweq} as
\begin{align}\label{eq:RGEeffpot}
    \partial_t u = &-D
    u +  (D\!-\!2\!+\!\eta_\phi)\bar{\rho}_\phi
    u^{(1,0)}  + (D\!-\!2\!+\!\eta_\chi)\bar{\rho}_\chi u^{(0,1)} \nonumber\\ 
    &+I_R^D(\omega_\chi, \omega_\phi,\omega_{\phi\chi})S_\phi(\tau)+(N\!-\!1)  I_G^D(u^{(1,0)})S_\phi(\tau)\nonumber\\
    & + I_R^D(\omega_\phi, \omega_\chi,\omega_{\phi\chi})S_\chi(\tau)\,.
\end{align}
Here the $S_i(\tau)=s_0^D(\tau) - \hat{s}_0^D(\tau) \frac{\eta_i}{D+2}$ with $i\in\{\phi, \chi\}$ depend on the reduced temperature $\tau = 2 \pi T / k$ and they contain all finite temperature effects through the thermal factors $s_0^D$ and $\hat{s}_0^D$, resulting from the Matsubara summation in Eq.~\eqref{eq:floweq}.
Note that for the linear regulator, finite temperature fluctuations completely factorize from zero temperature fluctuations written in terms of threshold functions $I_{R,G}^D$, which depend on the bosonic masses $\omega_\phi = u_k^{(1,0)} + 2 \bar{\rho}_\phi u_k^{(2,0)}$, $\omega_\chi = u_k^{(0,1)} + 2 \bar{\rho}_\chi u_k^{(0,2)}$, and $\omega_{\phi\chi}^2 = 4 \bar{\rho}_\phi \bar{\rho}_\chi (u_k^{(1,1)})^2$ with $u_k^{(l,m)} = \partial^l_{\rho_\phi}\partial^m_{\rho_\chi} u_k$. 

The first line in Eq.~\eqref{eq:RGEeffpot} originates from canonical dimensionality and field renormalizations. 
The second line arises from fluctuations of the radial mode and $N-1$~Goldstone modes in the $\mathrm{O}(N)$ sector and the third line contains the radial mode from the $\mathbb{Z}_2$ sector. 
Expressions for the anomalous dimensions $\eta_{\phi,\chi} = -\partial_t \log Z_{\phi, \chi}$, the explicit form of the thermal factors, and threshold functions are provided in~\cite{suppl}.

Here, we expand the effective potential $U_k[\rho_\phi,\rho_\chi]$ locally in the space of invariants around a (running) minimum~$(\rho_{\phi,0},\rho_{\chi,0})$. 
The expansion is done up to a finite power~$n$ of the invariants, and we refer to such a truncation as LPA$n$ if $Z_{\phi, \chi} = \mathrm{const.}$ and LPA${}'n$ if $Z_{\phi, \chi}$ is scale dependent. 
The position of the minimum corresponds to four different regimes, in which the system can be at RG scale~$k$: (1)~Both fields have a non-vanishing vacuum expectation value (vev) $v_{\phi,\chi}$ and the minimum lies at $(\kappa_\phi,\kappa_\chi)$ with $\kappa_{\phi,\chi} = v_{\phi,\chi}^2/2$ (SSB-SSB), i.e.,  $\mathrm{O}(N) \times \mathbb{Z}_2$ is broken down to $\mathrm{O}(N-1)$. 
(2)~\& (3)~Only one of the fields has a non-vanishing vev, i.e. $v_\phi > 0$ and $v_\chi = 0$ (SSB-SYM) or vice versa (SYM-SSB). 
In that case, the symmetry is either reduced to $\mathrm{O}(N-1) \times \mathbb{Z}_2$ or to $\mathrm{O}(N)$. 
(4)~Both fields have a vanishing vev (SYM-SYM), and the global symmetry is preserved.
The expansion coefficients of $U_k$ correspond to running couplings, e.g., the quartics $\lambda_{i}, i\in \{\phi,\chi,\phi\chi\}$, and their scale dependencies, i.e., $\beta$~functions, directly follow from Eq.~\eqref{eq:RGEeffpot}, cf.~\cite{suppl}. Below, dimensionless quantities are distinguished from their dimensionful counter-part by a bar on top.
 
Ultimately, the IR behaviour is extracted by solving the flow numerically towards $k\to 0$. 
If the system remains in the SSB regime in one (or both) scalar sectors, the corresponding vev fulfills $\lim_{k \to 0} v > 0$ and symmetry is broken in the thermodynamic state. Symmetry restoration is signaled by a vanishing vev as $k \to 0$.

\begin{figure}
    \centering
    \includegraphics[scale=0.75]{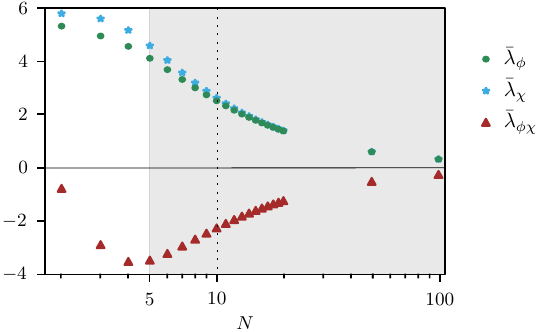}
    \caption{\textbf{BFP couplings for different~$N$.} Quartics $\bar{\lambda}_\phi$, $\bar{\lambda}_\chi$, $\bar{\lambda}_{\phi \chi}$ at the BFP as a function of $N$ in LPA$'6$. The grey-shaded area marks the points where $N |\bar{\lambda}_{\phi \chi}| > 3 \bar{\lambda}_\chi$, cf. Eq.~\eqref{eq:massflow}f.} 
    \label{fig:BFPquartics}
\end{figure}

\textit{Biconical fixed point. ---} 
Below $D=4$ dimensions, our model in Eq.~\eqref{action} is known to have several non-trivial RG fixed points~\cite{Fisher:1974,Vicari:2003,Eichhorn:2013mcl}.  
Here, we focus on the biconical fixed-point (BFP) in $D=2+1$ at $T=0$. 
We show the BFP coordinates of the quartic couplings with respect to $N$ within LPA${}'6$ in Fig.~\ref{fig:BFPquartics}. 
Notably, the $\bar{\lambda}_{\phi \chi}$ are negative for all $N \geq 2$, while the corresponding potentials are bounded from below. In~\cite{suppl}, we discuss the convergence of the LPA in detail.
%

\begin{figure}
    \centering
    \includegraphics[scale=0.85]{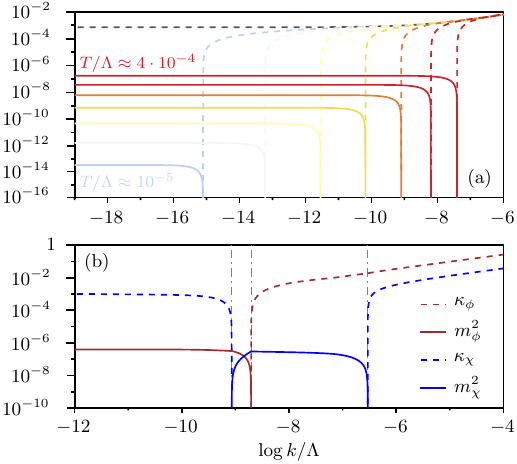}
    \caption{\textbf{Inverted SSB and CHMW from RG flows} for $N=100$. Panel~(a): Manifestation of CHMW for the $\mathrm{O}(N)$ sector. We show the flow of $\kappa_\phi$ in the SSB regime (dashed) and $m_\phi^2$ in the SYM regime (solid). We start the flow at at $\bar{\kappa}_\chi^\mathrm{UV} - \bar{\kappa}_\chi^\mathrm{BFP} = -0.005$ at zero temperature (dark gray curve) and increasing finite temperatures, as indicated by the colors. It is evident that CHMW is fulfilled, as the $\mathrm{O}(N)$ symmetry is always restored in the IR when $T>0$.
    Panel~(b): We show an exemplary flow for $\bar{\kappa}_\chi^\mathrm{UV} < \bar{\kappa}_\chi^\mathrm{BFP}$ above the phase boundary in Fig.~\ref{fig:pdN100}. The grey vertical lines mark the RG time at which the system changes its regime. The vertical axes are given in units of $\Lambda$.}
    \label{fig:mwh}
\end{figure}

\textit{Mechanism for high-$T$ SSB. ---} We discuss how high-$T$ SSB emerges from functional methods. For simplicity, we restrict to LPA4, but the reasoning holds beyond truncations.
We start by showing how CHMW manifests itself in the flow. 
To that end, we assume that at some scale $k$ the system is in a SSB regime with $\kappa_\phi > 0$. 
From Eq.~\eqref{eq:RGEeffpot}, the flow of $\kappa_\phi$ high $T$ or small~$k$ is given by~\cite{suppl}
\begin{align}\label{eq:flowcondphi}
    \partial_t \kappa_\phi = k \partial_k \kappa_\phi \approx a_d T (N-1)k^{d-2} + \mathcal{O}(k^{D+1})
\end{align}
with positive constant $a_d$. If $N = 1$ or $d > 2$, the flow gets arbitrarily slow towards the IR, and thus $\kappa_\phi > 0$ is possible for $k\to 0$. 
However, if $N > 1$ and $d \leq 2$, $\partial_k \kappa_\phi$ gets large towards the IR and eventually drives $\kappa_\phi$ to zero and into SYM. We explicitly show this in~\cite{suppl}.
This means that for $d \leq 2$ it is not possible to maintain $\kappa_\phi > 0$ at $T > 0$ towards the IR if $N > 1$, which is precisely the statement of CHMW.
We show the numerical manifestation of this 
in Fig.~\ref{fig:mwh}.

We now turn to the $\mathbb{Z}_2$ sector in $D=2+1$ where the first term in Eq.~\eqref{eq:flowcondphi} is absent in $\partial_t \kappa_\chi$.
Thus the fate of $\mathbb{Z}_2$ symmetry-breaking solely depends on the values of the couplings.
To understand how inverted SSB of $\mathbb{Z}_2$ is possible, it is sufficient to consider the case where $\mathrm{O}(N)$ symmetry is already restored and $m_\chi^2 > 0$ at some $k$, as a vanishing vev corresponds to a positive mass term~\cite{suppl}.
The flow of $m_\chi^2$ for high $T$ is given by
\begin{align}\label{eq:massflow}
    \partial_t m_\chi^2 &=  - \frac{k^4 a_D T}{3 \pi^2} \Big( \frac{3\lambda_\chi}{(k^2+m_\chi^2)^2} +\frac{ N \lambda_{\phi\chi}}{(k^2+m_\phi^2)^2} \Big)\,.
\end{align}
The first term on the right-hand side $\propto \lambda_\chi$ is always negative as we demand $\lambda_\chi >0$ for boundedness of $U_k$. 
Therefore, 
if the coupling $\lambda_{\phi\chi}$ between the two scalar sectors was absent, the IR flow would always lead to an increase of $m_\chi^2$ and the system could never leave the SYM regime. However, $\lambda_{\phi \chi} < 0$ is allowed and yields fluctuations decreasing $m_\chi^2$ towards the IR~\footnote{We note that this is different for fermionic thermal factors, which vanish for $\tau \geq 2$ due to the absence of a zeroth Matsubara mode, cf.~\cite{Scherer:2013many, Tolosa:2024}. This therefore excludes the possibility of spontaneous symmetry-breaking at high temperatures through a Yukawa interaction.}. 
Qualitatively, a criterion for inverted SSB can be deduced by neglecting the non-perturbative denominators in Eq.~\eqref{eq:massflow}: It is only possible to flip the sign of $\partial_t m_\chi^2$ and make $m_\chi^2$ decrease towards the IR, if at some scale $N |\lambda_{\phi \chi}| > 3 \lambda_\chi$, see Fig.~\ref{fig:BFPquartics}.
Whether a transition into the $\mathbb{Z}_2$ SSB regime actually happens needs to be determined by numerical integration of the RG flow towards the IR, see below.

\textit{Persistent SSB. ---} 
Let us collect what is necessary for persistent SSB: 
We need $\lambda_{\phi \chi} < 0$ during the flow and $N$ has to be sufficiently large, such that $\partial_t m_\chi^2 \supset - N \lambda_{\phi \chi}$ dominates the mass-increasing contributions from the self-coupling $\lambda_\chi$.
In addition, at $T = 0$, the $\mathbb{Z}_2$~symmetric and broken phase have to be separated by a QCP, defining a CFT. 
The natural candidate fulfilling these conditions is the BFP, cf. Fig.~\ref{fig:BFPquartics}.
To solve the RG flow in its vicinity we use $\bar{\kappa}_\chi$ as a tuning parameter. 
Specifically, we start the flow at a UV scale $\Lambda$ defining the units and by setting all couplings to their respective BFP values. 
We then slightly perturb the UV value $\bar{\kappa}_\chi^\mathrm{UV}$ from its fixed-point value~$\bar{\kappa}_\chi^\mathrm{BFP}$ and integrate the flow to the IR at fixed~$T$.
Finally, we examine whether $\mathbb{Z}_2$ symmetry is broken or not.
For explicit calculations, we employ LPA${}'6$, which we establish in~\cite{suppl} to provide well-converged results in the considered range of $N$.
LPA${}'6$ also turns out to yield the numerically most stable results, at least within finite orders of the LPA, cf.~\cite{suppl}.


\begin{figure}
    \centering
    \includegraphics[scale=0.77]{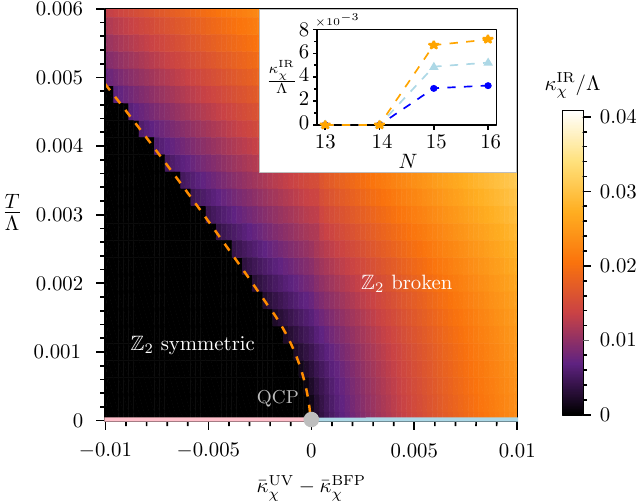}
    \caption{\textbf{Phase diagram} for $N=100$ in LPA${}'6$. For $\bar{\kappa}_\chi^\mathrm{UV} < \bar{\kappa}_\chi^\mathrm{BFP}$, we find inverted SSB of $\mathbb{Z}_2$ above a critical temperature, while $\mathbb{Z}_2$ is spontaneously broken at all temperatures for $\bar{\kappa}_\chi^\mathrm{UV} > \bar{\kappa}_\chi^\mathrm{BFP}$. Additionally, the $\mathrm{O}(N)$ symmetry is broken on both sides of the QCP at $T = 0$. The phase boundary separating $\mathbb{Z}_2$ symmetric and broken phases, as indicated by the orange dashed line, is thus bent to the left, which is a direct manifestation of persistent SSB. The coloured thick lines left and right of the QCP indicate the $\mathrm{O}(N-1) \times \mathbb{Z}_2$ and $\mathrm{O}(N-1)$ regimes at $T = 0$, respectively.
    Inset: Critical $N$ in LPA${}'6$. We show $\kappa_\chi^\mathrm{IR}$ for $\bar{\kappa}_\chi^\mathrm{UV} = \bar{\kappa}_\chi^\mathrm{BFP}$ and increasing temperature as indicated by the symbols, i.e., $T/\Lambda=0.01$ (stars), $T/\Lambda=0.0075$ (triangles), $T/\Lambda=0.005$ (circles). The figure suggests that persistent symmetry breaking occurs for $N \geq N_c = 15$. LPA6 yields the same estimate.} 
    \label{fig:pdN100}
\end{figure}

\textit{Large $N$ and critical $N$. ---} At large finite values of $N$, we can expect results that are well-converged within the LPA${}'$, see~\cite{suppl} for more details. 
We thus start our discussion for $N = 100$ and show the resulting $\bar{\kappa}_\chi^\mathrm{UV}$-$T$ phase diagram in Fig.~\ref{fig:pdN100}.
At $T=0$, the BFP separates two phases: for $\bar{\kappa}_\chi^\mathrm{UV} > \bar{\kappa}_\chi^\mathrm{BFP}$, the symmetry is broken down to $\mathrm{O}(N-1)$, while it is broken down to $\mathrm{O}(N-1) \times \mathbb{Z}_2$ for $\bar{\kappa}_\chi^\mathrm{UV} < \bar{\kappa}_\chi^\mathrm{BFP}$. 
These phases are indicated by the thick lines on the $T=0$ axis.
When turning on the temperature, we find that the symmetry in the $\mathrm{O}(N)$ sector is always restored in the IR, owing to CHMW, cf. Fig.~\ref{fig:mwh}. 
For $\bar{\kappa}_\chi^\mathrm{UV} < \bar{\kappa}_\chi^\mathrm{BFP}$, we find a transition from a small-$T$ $\mathbb{Z}_2$ symmetric to a high-$T$ $ \mathbb{Z}_2$ broken phase, i.e., inverted SSB. 
For $\bar{\kappa}_\chi^\mathrm{UV} > \bar{\kappa}_\chi^\mathrm{BFP}$, persistent breaking of  $\mathrm{O}(N)\times \mathbb{Z}_2 \to \mathrm{O}(N)$ occurs. 

We show an exemplary flow for $\bar{\kappa}_\chi^\mathrm{UV} < \bar{\kappa}_\chi^\mathrm{BFP}$ above the inverted finite-$T$ transition in Fig.~\ref{fig:mwh}. 
Starting near the BFP within the SSB-SSB phase, fluctuations first lead to a decrease of $\kappa_\phi$ and $\kappa_\chi$, and at some intermediate $k$ the global symmetry is fully restored. However, as soon as the $\mathrm{O}(N)$ symmetry is restored, $m_\chi^2$ is monotonically decreasing and the system enters the SYM-SSB phase, cf. the discussion above. 
Below the critical temperature, $\kappa_\chi$ again goes to zero at intermediate RG scales and $\mathbb{Z}_2$ symmetry is restored, while $\kappa_\chi > 0$ down to the IR above the critical temperature and $\mathbb{Z}_2$ is broken.

Persistent SSB manifests itself through a phase boundary being bent to the left, since then, right above the QCP, we find the ordered state and any finite temperature is equivalent to any other~\cite{PhysRevLett.125.131603, Chai:2020zgq}, see also Fig.~\ref{fig:pdN100}. 
This implies that if a critical value $N_c$ of $N$ exists, then the phase boundary is bent to the right for $N < N_c$ and it gets vertical precisely at $N_c$. 
Therefore, our strategy for determining $N_c$ is to set $\bar{\kappa}_\chi^\mathrm{UV} = \bar{\kappa}_\chi^\mathrm{BFP}(N)$ for some choice of $N$ and at a fixed finite $T$ and to then examine the value of $\kappa_\chi^\mathrm{IR}$. 
If $N < N_c$, then $\kappa_\chi^\mathrm{IR} = 0$ for all $T > 0$, while, if $N > N_c$, $\kappa_\chi^\mathrm{IR} > 0$ for all $T > 0$. 
In the inset of Fig.~\ref{fig:pdN100}, we show $\kappa_\chi^\mathrm{IR}$ for $13 \leq N \leq 16$ for various finite temperatures.  
Restricting to integer $N$, we conclude that $N_c \approx 15$.
This can be compared with perturbative results. 
Naively extrapolating the 1-loop result of~\cite{PhysRevLett.125.131603, Chai:2020zgq}, we obtain $N_c=17$. 
Alternatively, $N_c$ may be inferred from non-local models~\cite{Chai:2021djc}, yielding $N_c=17$ at the leading order~\footnote{This works as follows: Taking the scalars to have scaling dimension $\Delta=\frac{3-\delta}{4}$, one can perform reliable perturbative calculations at $\delta\rightarrow 0$ and get an $N_c$ which depends on the perturbation parameter $\delta$.
To extract information for the short-range model, one needs to impose the local conformal Ward identity $\Delta_{T^{\mu\nu}}=3$ as recently discussed in Ref.~\cite{Rong:2024vxo}. 
This allows us to solve for the $\delta_{\rm local}$ that corresponds to the local CFT}.


\textit{Conclusion. ---}  We addressed several open questions on biconical $\mathrm{O}(N) \times \mathbb{Z}_2$ models in the context of persistent symmetry breaking.
Concretely, using functional methods, we showed directly in $D=2+1$ that the models possess a UV completion allowing SSB at all temperatures.
Desisting from expansion in fractional dimension we avoid the issue of unitarity violation~\cite{PhysRevD.93.125025} and we do not introduce non-localities~\cite{Chai:2021djc}.
Moreover, we transparently exhibit the mechanisms leading to inverted SSB as well as the fulfillment of CHMW  wherever applicable and further resolve the full finite-$T$ phase diagrams of the model.
We determined the critical value of $N$ above which persistent symmetry-breaking occurs to be $N_c = 15$, which is close to estimates that can be drawn from related non-local models or the $4 - \epsilon$ expansion.

For the future, it will be interesting to study whether the value of $N_c$ can be modified, e.g., by coupling to additional matter, enlarging the discrete symmetry to $\mathbb{Z}_m$ with $m > 2$, or by considering more exotic discrete symmetries~\cite{Liendo:2022bmv}.


\begin{acknowledgments}
\textit{Acknowledgements. ---} 
We thank Laura Classen, Astrid Eichhorn, and Holger Gies for comments on the manuscript  and Elio K\"onig and  Fabian Rennecke for discussions. 
BH and MMS are supported by the Mercator Research Center Ruhr under Project No. Ko-2022-0012.
MMS acknowledges funding from the Deutsche Forschungsgemeinschaft (DFG, German Research Foundation) within Project-ID 277146847, SFB 1238 (project C02), and the DFG Heisenberg programme (Project-ID 452976698). This research was supported in part by grant NSF PHY-2309135 to the Kavli Institute for Theoretical Physics (KITP).
\end{acknowledgments}

\bibliographystyle{apsrev4-1}
\bibliography{allTSSB}


\clearpage
\pagebreak
\appendix
\widetext
\begin{center}
\textbf{\large Supplemental Material:\\[5pt]
UV complete local field theory of persistent symmetry breaking in 2+1 dimensions
}
\end{center}
\setcounter{equation}{0}
\setcounter{figure}{0}
\setcounter{table}{0}

\section{Flow Equations of $\mathrm{O}(N)\times \mathrm{O}(M)$ Models at Finite Temperature}
\label{app:proj}

We employ the extended local potential approximation for the effective average action 
\begin{equation}
    \Gamma_k[\phi, \chi] = \int d^Dx \bigg[ \frac{Z_\phi}{2}(\partial \phi)^2 + \frac{Z_\chi}{2} (\partial \chi)^2 + U_k[\rho_\phi,\rho_\chi] \bigg],
\end{equation}
with uniform field renormalizations $Z_{\phi,\chi}$. 
The FRG flow equation for the dimensionless effective potential, defined by $u_k = k^{-D} U_k$ as a functional of the dimensionless invariants $\bar{\rho}_{\phi,\chi} = Z_{\phi,\chi}k^{2-D}\rho_{\phi,\chi}$, is then obtained by projecting $\partial_t \Gamma_k$ onto constant field-configurations and reads
\begin{align}\label{eq:RGEeffpotfull}
    \partial_t u_k = &-D
    u_k +  (D - 2 + \eta_\phi)\bar{\rho}_\phi
    u_k^{(1,0)}  + (D - 2 + \eta_\chi)\bar{\rho}_\chi u_k^{(0,1)} \nonumber\\[8pt]
    &+ \left[
    I_R^D(\omega_\chi, \omega_\phi,\omega_{\phi\chi}) + (N - 1)  I_G^D(u_k^{(1,0)})  \right]S_\phi(\tau) \nonumber\\[8pt]
    & + \left[ 
    I_R^D(\omega_\phi, \omega_\chi,\omega_{\phi\chi}) + (M - 1)  I_G^D(u_k^{(0,1)}) \right]S_\chi(\tau),
\end{align}
where we defined $u_k^{(l,m)} = \partial^l_{\rho_\phi} \partial^m_{\rho_\chi} u_k$. 
Further, $S_i(\tau)=s_0^D(\tau) - \hat{s}_0^D(\tau) \frac{\eta_i}{D+2}$ with $\eta_i = -\partial_t \log Z_i$ for $i\in\{\phi, \chi\}$ contains all finite temperature effects, which are fully encoded in the thermal factors $s_0^D$ and $\hat{s}_0^D$ as a function of the reduced temperature $\tau = 2 \pi T / k$. 

The first line in Eq.~\eqref{eq:RGEeffpotfull} arises from canonical dimensionality and non-trivial field renormalizations, while the second and third line originate from fluctuations of Goldstone and radial modes of the $\mathrm{O}(N)$ and $\mathrm{O}(M)$ sector, respectively. 
In the above Eq.~\eqref{eq:RGEeffpotfull}, the threshold functions have been evaluated for a linear (optimized) regulator~\cite{PhysRevD.64.105007,Litim:2001fd}, which regularizes ``covariantly" in (bosonic) Matsubara frequencies $\omega_n=2\pi n T, n\in \mathbb{Z}$, and spatial momenta $\vec{p}$,
\begin{align}
    R_k^i = Z_i ((2\pi n T)^2 + \vec{p}^2) r\left(\frac{(2\pi n T)^2 + \vec{p}^2}{k^2}\right)\,,\ \text{where}\ r(y) = \left(\frac{1}{y} - 1\right) \theta(1-y)\,,\quad i\in\{\phi,\chi\}\,.
\end{align}
Explicit expressions for the threshold functions for that choice of regulator are given in the next section.
A convenient feature of the linear regulator is that thermal fluctuations completely factorize from zero temperature fluctuations. Eq.~\eqref{eq:RGEeffpotfull} can be used as the starting point for solutions of the full potential in the space of invariants, cf., e.g., Refs.~\cite{Borchardt:2016,Grossi:2019urj,Ihssen:2022xkr}.

As discussed in the main text, we expand the effective potential locally around a minimum $(\rho_{\phi,0},\rho_{\chi,0})$. The physical situations where both fields (SSB-SSB), one of the fields (SYM-SSB) or (SSB-SYM) or none of the fields (SYM-SYM) obtain a non-vanishing vacuum expectation value at RG scale $k$ admit a suitable local expansion and read in terms of dimensionless quantities
\begin{align}
    \text{SYM-SYM}&: \quad  u_k[\bar{\rho}_\phi,\bar{\rho}_\chi] = \bar{m}_\phi^2 \bar{\rho}_\phi + \bar{m}_\chi^2 \bar{\rho}_\chi + \sum_{n+m\geq 2} \frac{\bar{\lambda}_{n,m}}{n! m!} \bar{\rho}_\phi^n \bar{\rho}_\chi^m \label{eq:localu1}\\[8pt]
    \text{SYM-SSB}&: \quad  u_k[\bar{\rho}_\phi,\bar{\rho}_\chi] = \bar{m}_\phi^2 \bar{\rho}_\phi + \sum_{n+m\geq 2} \frac{\bar{\lambda}_{n,m}}{n! m!} \bar{\rho}_\phi^n (\bar{\rho}_\chi-\bar{\kappa}_\chi)^m \\[8pt]
    \text{SSB-SYM}&:  \quad  u_k[\bar{\rho}_\phi,\bar{\rho}_\chi] = \bar{m}_\chi^2 \bar{\rho}_\chi + \sum_{n+m\geq 2} \frac{\bar{\lambda}_{n,m}}{n! m!} (\bar{\rho}_\phi-\bar{\kappa}_\phi)^n \bar{\rho}_\chi^m\\[8pt]
    \text{SSB-SSB}&: \quad  u_k[\bar{\rho}_\phi,\bar{\rho}_\chi] = \sum_{n+m\geq 2} \frac{\bar{\lambda}_{n,m}}{n! m!} (\bar{\rho}_\phi-\bar{\kappa}_\phi)^n (\bar{\rho}_\chi-\bar{\kappa}_\chi)^m\label{eq:localu4}
\end{align}
with $\bar{\kappa}_{\phi,\chi} > 0$ and $\bar{m}_{\phi,\chi}^2 > 0$. Note that the expansion in  Eq.~\eqref{eq:localu4} is well-defined only if $\Delta := \bar{\lambda}_{1,0}\bar{\lambda}_{0,1} - \bar{\lambda}_{1,1}^2$ is positive.
\clearpage

The RG equations of the dimensionless scalar couplings can be then obtained self-consistently from Eq.~\eqref{eq:RGEeffpotfull} by the following projection prescriptions:
\begin{enumerate}
    \item SYM-SYM:
    \begin{align}
        \partial_t \bar{m}_{\phi}^2 &= \partial_t u_k^{(1,0)}\Big|_{\substack{\bar{\rho}_\phi=0 \\ \bar{\rho}_\chi=0}},\quad
        \partial_t \bar{m}_{\chi}^2 = \partial_t u_k^{(0,1)}\Big|_{\substack{\bar{\rho}_\phi=0 \\ \bar{\rho}_\chi=0}},\quad
        \partial_t \bar{\lambda}_{n,m} = \partial_t u_k^{(n,m)}\Big|_{\substack{\bar{\rho}_\phi=0 \\ \bar{\rho}_\chi=0}}.
    \end{align}
    \item SSB-SYM:
    \begin{align}
        \partial_t \bar{\kappa}_\phi&=- \frac{\partial_t u_k^{(1,0)}}    {u_k^{(2,0)}}\Big|    _{\substack{\bar{\rho}_\phi=\bar{\kappa}_\phi \\    \bar{\rho}_\chi=0}},\ 
        \partial_t \bar{m}_\chi^2=\left(\partial_t u_k^{(0,1)}+u_k^{(1,1)} \partial_t \bar{\kappa}_\phi\right)\Big|_{\substack{\bar{\rho}_\phi=\bar{\kappa}_\phi \\ \bar{\rho}_\chi=0}}, \ 
        \partial_t \bar{\lambda}_{l, m} =\left(\partial_t u_k^{(l, m)}+u_k^{(l+1, m)} \partial_t \bar{\kappa}_\phi \right)\Big|_{\substack{\bar{\rho}_\phi=\bar{\kappa}_\phi \\ \bar{\rho}_\chi=0}}.
    \end{align}
    \item SYM-SSB: same as SSB-SYM with $\phi \leftrightarrow \chi$.
    \item SSB-SSB:
    \begin{align}
        \partial_t \bar{\kappa}_\phi&=\frac{u_k^{(0,2)} \partial_t u_k^{(1,0)}-u_k^{(1,1)} \partial_t u_k^{(0,1)}}{\left(u_k^{(1,1)}\right)^2-u_k^{(2,0)} u_k^{(0,2)}}\Bigg|_{\substack{\bar{\rho}_\phi=\bar{\kappa}_\phi\\ \bar{\rho}_\chi=\bar{\kappa}_\chi}},\quad
        \partial_t \bar{\kappa}_\chi=\frac{u_k^{(2,0)} \partial_t u_k^{(0,1)}-u_k^{(1,1)} \partial_t u_k^{(1,0)}}{\left(u_k^{(1,1)}\right)^2-u_k^{(2,0)} u_k^{(0,2)}}\Bigg|_{\substack{\bar{\rho}_\phi=\bar{\kappa}_\phi \\ \bar{\rho}_\chi=\bar{\kappa}_\chi}} ,\\[8pt]
        \partial_t \lambda_{l, m} &=\left(\partial_t u_k^{(l, m)}+u_k^{(l+1, m)} \partial_t \bar{\kappa}_\phi+u_k^{(l, m+1)} \partial_t \bar{\kappa}_\chi\right)\Big|_{\substack{\bar{\rho}_\phi=\bar{\kappa}_\phi \\ \bar{\rho}_\chi=\bar{\kappa}_\chi}}.
    \end{align}
\end{enumerate}
To complete the set ot RG equations, we determine the anomalous dimensions $\eta_{\phi,\chi} = - \partial_t \log\, Z_{\phi, \chi}$ using a background-field expansion {$\phi^a = \phi_0^a \delta^{a,1} + \xi_\phi^a$} and {$\chi^a = \chi_0^a \delta^{a,1} + \xi_\chi^a$}. The anomalous dimensions can then be extracted from the flow equations of the field-renormalizations  
\begin{align}
    \partial_t Z_{\phi,\chi} \delta^{(D)}(0) &= \frac{\delta^{\mu \nu}}{2D} \partial_\mu \partial_\nu \frac{\delta^2}{\delta \xi^a_{\phi,\chi}(q) \delta \xi^a_{\phi,\chi}(-q)} \partial_t \Gamma_k \bigg|_{\substack{\xi_\phi = \xi_\chi = 0 \\ q^2 = 0}},
\end{align}
with $a \in \{2,..,N\}$ in order to project onto one of the Goldstone modes. It has been shown that, even in the case of $\mathrm{O}(1) \cong \mathbb{Z}_2$, projecting onto one of the Goldstone modes and then taking the limit $M \to 1$ yields better results than by projecting onto the radial mode, cf. Refs.~\cite{Berges:2002,PhysRevB.93.125119}. The evaluation of the right hand side is done by expanding the regularized field-dependent propagator in a constant and fluctuating part, i.e. $\Gamma^{(2)}[\Phi] = \Gamma^{(2)}[\Phi_0] + \Delta \Gamma^{(2)}[\Phi_0,\xi]$. The flow equation of the flowing action can then be expanded in powers of the fluctuating field $\xi$,
\begin{align}
    \partial_t \Gamma_k[\Phi] &= \frac{1}{2} \tilde{\partial_t} \mathrm{Tr} \log\left( \Gamma^{(2)}_k[\Phi] + R_k\right) \\[10pt]
    &= \frac{1}{2} \tilde{\partial_t} \mathrm{Tr} \log\left( \Gamma^{(2)}_k[\Phi_0] + R_k\right)  + \frac{1}{2} \tilde{\partial_t} \mathrm{Tr} \sum_{n=1}^\infty \frac{(-1)^{n+1}}{n} \left[ (\Gamma^{(2)}_k[\Phi_0] + R_k)^{-1} \Delta \Gamma^{(2)}[\Phi_0,\xi] \right]^n.
\end{align}
In the above, the operator $\tilde{\partial_t}$ acts only on the $k$-dependence of $R_k$. The only $q$-dependent contribution comes from the $n=2$ term, which yields 
\begin{align}
    \partial_t Z_{\phi,\chi} \delta^{(D)}(0) &= -\frac{\delta^{\mu \nu}}{8D} \partial_\mu \partial_\nu \mathrm{Tr} \left[ G_{k,0} \frac{\delta \Delta \Gamma_k^{(2)}}{\delta \xi_{\phi,\chi}^a(q)} G_{k,0} \frac{\delta \Delta \Gamma_k^{(2)}}{\delta \xi_{\phi,\chi}^a(-q)} \right] \bigg|_{\substack{\xi_\phi = \xi_\chi = 0 \\ q^2 = 0}}
\end{align}
with $G_{k,0}[\Phi_0] = (\Gamma^{(2)}_k[\Phi_0] + R_k)^{-1}$. We find, independent of $N$ and $M$,
\begin{align}\label{eq:anomdims}
    \eta_\phi &=  2 \sqrt{2 \bar{\rho}_\chi} u^{(1,1)}_k s^D_0(\tau) m^D_{R_\phi G_\phi}(\sqrt{2 \bar{\rho}_\chi}u^{(1,1)}_k,\sqrt{2 \bar{\rho}_\phi}u_k^{(2,0)};\omega_\chi,\omega_\phi,\omega_{\phi \chi},u_k^{(1,0)})\notag \\ 
     & \hspace{1.2cm} + 2\sqrt{2 \bar{\rho}_\phi} u_k^{(2,0)} s^D_0(\tau) m^D_{R_\chi G_\phi}(\sqrt{2 \bar{\rho}_\phi}u_k^{(2,0)},\sqrt{2 \bar{\rho}_\chi}u_k^{(1,1)};\omega_\phi,\omega_\chi,\omega_{\phi \chi},u_k^{(1,0)}), \\[8pt]
    \eta_\chi &= 2 \sqrt{2 \bar{\rho}_\phi} u_k^{(1,1)} s^D_0(\tau) m^D_{R_\chi G_\chi}(\sqrt{2 \bar{\rho}_\phi}u_k^{(1,1)},\sqrt{2 \bar{\rho}_\chi}u_k^{(0,2)};\omega_\phi,\omega_\chi,\omega_{\phi \chi},u_k^{(0,1)})\notag\\ 
     & \hspace{1.2cm} + 2\sqrt{2 \bar{\rho}_\chi} u_k^{(0,2)} s^D_0(\tau) m^D_{R_\phi G_\chi}(\sqrt{2 \bar{\rho}_\chi}u_k^{(0,2)},\sqrt{2 \bar{\rho}_\phi}u_k^{(1,1)};\omega_\chi,\omega_\phi,\omega_{\phi \chi},u_k^{(0,1)}).
\end{align}


\section{Threshold Functions and Thermal Factors}
\label{app:thf}

The threshold functions appearing in the flow equation~\eqref{eq:RGEeffpotfull} evaluated for the linear regulator read
\begin{align}
    I_R^D(x,y,z) &= \frac{4v_D}{D}\frac{1+x}{(1+x)(1+y) - z^2}, \quad
    I_G^D(x) = \frac{4v_D}{D}\frac{1}{1+x}, \\
    m^D_{R_\varphi G_\vartheta} 
    &= \frac{4 v_D}{D} \bigg[ \frac{1+\omega_\theta}{(1+\omega_\vartheta)^2}\frac{(1+\omega_\theta)v_1 - \omega_{\varphi \theta}v_2}{((1+\omega_\varphi)(1+\omega_\theta) - \omega_{\varphi\theta}^2)^2}  - \frac{\omega_{\varphi\theta}}{(1+\omega_\vartheta)^2}\frac{(1+\omega_\varphi)v_2 - \omega_{\varphi \theta}v_1}{((1+\omega_\varphi)(1+\omega_\theta) - \omega_{\varphi\theta}^2)^2}\bigg],
\end{align}
with $m^D_{R_\varphi G_\vartheta}=m^D_{R_\varphi G_\vartheta}(v_1,v_2;\omega_\varphi,\omega_\theta,\omega_{\varphi \theta},\omega_\vartheta)$,  $v_D^{-1} = 2^{D+1} \pi^{D/2} \Gamma(D/2)$, and the bosonic masses 
\begin{align}
    \omega_\phi &= u_k^{(1,0)} + 2 \bar{\rho}_\phi u_k^{(2,0)}, \quad
    \omega_\chi = u_k^{(0,1)} + 2 \bar{\rho}_\chi u_k^{(0,2)}, \quad
    \omega_{\phi\chi}^2 = 4 \bar{\rho}_\phi \bar{\rho}_\chi (u_k^{(1,1)})^2.
\end{align}
The thermal factors as a function of the reduced temperature $\tau = 2 \pi T/k$ read
\begin{align}
    s_0^D(\tau) &= \frac{v_{D-1}}{v_D} \frac{D}{D-1} \frac{\tau}{2 \pi} \sum_{n \in \mathbb{Z}} \Theta(1-n^2\tau^2)(1-n^2\tau^2)^{\frac{D-1}{2}}, \\
    \hat{s}_0^D(\tau) &= \frac{v_{D-1}}{v_D} \frac{D}{D-1} \frac{D+2}{D+1} \frac{\tau}{2 \pi} \sum_{n \in \mathbb{Z}} \Theta(1-n^2\tau^2)(1-n^2\tau^2)^{\frac{D+1}{2}},
\end{align}
and are normalized in the sense that $\lim_{T \to 0} s_0^D(\tau) = \lim_{T \to 0} \hat{s}_0^D(\tau) = 1$.
In $D=2+1$, the Matsubara sums can be performed analytically and yield
\begin{align}
    s_0^D(\tau) &= \frac{v_{D-1}}{v_D} \frac{D}{D-1} \frac{\tau}{2 \pi} \frac{1}{3}\left(1 + 2s_B(\tau)\right) \left[ 3 - \tau^2 s_B(\tau)\left(1 + s_B(\tau)\right) \right], \\
    \hat{s}_0^D(\tau) &= \frac{v_{D-1}}{v_D} \frac{D}{D-1} \frac{D+2}{D+1} \frac{\tau}{2 \pi} \frac{1}{15} \left( 1 + 2 s_B(\tau) \right) \bigg[ 15 + \tau^2 s_B(\tau) \left( 1 + s_B(\tau) \right) \left( -10 - \tau^2 + 3\tau^2 s_B(\tau) \left( 1 + s_B(\tau) \right) \right) \bigg] 
\end{align}
where $s_B(\tau) = \floor*{\frac{1}{\tau}}$. For $\tau > 1$, $s_B(\tau) = 0$, and thus only the zeroth Matsubara mode survives at high temperatures or small scales.

\begin{figure}[t!]
    \centering
    \includegraphics[scale=0.25]{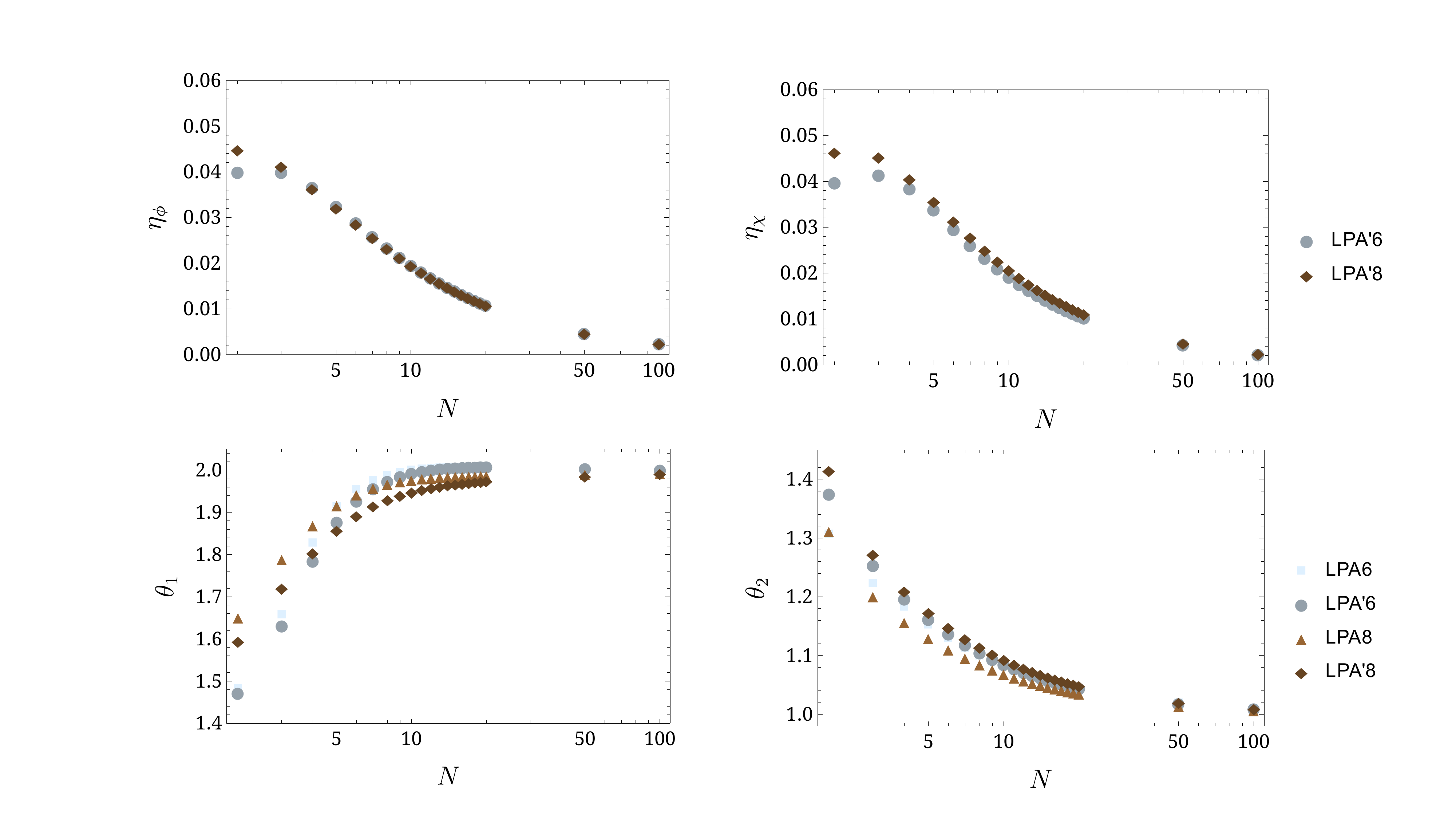}
    \caption{Anomalous dimensions $\eta_{\phi}$ and $\eta_{\chi}$ (top) and leading negative eigenvalues of the stability matrix (bottom) at the BFP within LPA${}'6$ and LPA${}'8$ for different values of $N$.}
    \label{fig:etasathetas}
\end{figure}

We note that there are ambiguities when continuing momentum integrals to $T>0$. At $T = 0$, consider for example an integral of the form $\int_q q^2 f(q^2)$. Because of rotational symmetry in spacetime, we have for any $\mu \in \{ 1,...,D \}$
\begin{equation}\label{eq:ambigcont}
    \int \frac{d^Dq}{(2 \pi)^D} q^2 f(q^2) = D \int \frac{d^Dq}{(2 \pi)^D} (q^\mu)^2 f(q^2) = \frac{D}{D-1} \int \frac{d^Dq}{(2 \pi)^D} \vec{q}^2 f(q^2).
\end{equation}
When continuing the theory to $T>0$, we introduce an anisotropy between space and time by compactifying the time domain. 
Therefore, the result for $T > 0$ depends on the choice made at $T = 0$, since the relation \eqref{eq:ambigcont} does not hold for arbitrary continuations to $T > 0$. Such an integral appears for example in the evaluation of the anomalous dimension, where, if we had continued the first expression in Eq.~\eqref{eq:ambigcont}, the thermal factor 
\begin{equation}
    t_0^D(\tau) = \frac{v_{D-1}}{v_D} \frac{\tau}{2\pi} \sum_n \Theta(1-n^2\tau^2)(1-n^2 \tau^2)^{\frac{D-3}{2}}
\end{equation}
would appear in Eq.~\eqref{eq:anomdims}, while when continuing the last expression in Eq.~\eqref{eq:ambigcont}, $s_0^D$ appears in Eq.~\eqref{eq:anomdims}. Further, we note that in $D=2+1$ and for $\tau > 0$, $s_0^D(\tau) = t_0^D(\tau) + \mathcal{O}(\tau^2)$ and $s_0^D(\tau) = t_0^D(\tau)$ for $\tau > 1$. We are not aware of any advantages or disadvantages that might accompany those choices and, since the results are independent of the precise continuation, we choose $s_0^D$ as this choice has been already successfully used before~\cite{Scherer:2013many}.


\subsection{Manifestation of Coleman-Hohenberg-Mermin-Wagner Theorem in Flow Equations}

We show how CHMW manifests itself in the flow equations. 
To that end, we assume that at some scale $k$ the system is in a SSB regime with $\kappa_\phi > 0$. 
Following from Eq.~\eqref{eq:RGEeffpotfull}, the flow of $\kappa_\phi$ is given by
\begin{align}
    \partial_t \kappa_\phi = \frac{4 v_D}{D} s_0^D(\tau) k^{D+2} \frac{N - 1}{k^4} + \mathcal{O}(k^{D+1}).
\end{align}
The first term arises from the $N-1$ Goldstone fluctuations in the $\mathrm{O}(N)$ sector. 
For high $T$ or small~$k$, we have ${2 \pi T / k > 1}$ and then $s_0^D(\tau)$ simplifies to ${s_0^D(\tau) = \frac{v_{d}}{v_D}\frac{D}{d} \frac{T}{k}}$, yielding
\begin{align}
    \partial_t \kappa_\phi = k \partial_k \kappa_\phi \approx a_d T (N-1)k^{d-2} + \mathcal{O}(k^{D+1})
\end{align}
with $a_d = 2 v_{d} / \pi d$. 
If $N = 1$ or $d > 2$, the flow gets arbitrarily slow towards the IR, and thus $\kappa_\phi > 0$ is possible for $k\to 0$. 
However, if $N > 1$ and $d \leq 2$, $\partial_k \kappa_\phi$ gets large towards the IR and eventually drives $\kappa_\phi$ to zero and into SYM. 
More precisely, for $d=2$ the first term dominates below a finite scale $k_i$, and the flow of $\kappa_\phi$ for $k < k_i$ is 
\begin{align}
    \kappa_\phi(k) \approx
       \kappa_\phi(k_i) - a_D T (N-1) \log k_i/k\,.
\end{align}
Therefore, $ \kappa_\phi = 0$ at a finite scale $k_0$ with $0 < k_0 < k_i$ and $k_0 =k_i \exp\, (-\kappa_\phi(k_i)/a_D (N-1) T)$, which holds for any positive values of $k_i$ and $\kappa_\phi(k_i)$,  and hence the system is in the symmetric regime. Analogously for $d=1$, we find for $k>k_i$ that 
\begin{align}
\kappa_\phi(k) = \kappa_\phi(k_i) - a_D T (N-1) (1/k - 1/k_i)
\end{align}
and hence the squared vev dies out at $k_0=k_i/[1+(k_i \kappa_\phi(k_i))/(a_D (N-1) T)]$.
This means that for $d \leq 2$ it is not possible to maintain $\kappa_\phi > 0$ at $T > 0$ towards the IR if $N > 1$, which is precisely the statement of CHMW.
We conclude by noting that the Mermin-Wagner theorem makes a statement about the unrenormalized vev $v_\phi = \sqrt{2 \kappa_\phi}$ and hence the inclusion of the anomalous dimension does not change the above arguments. As long as the field renormalization stays finite towards the IR, the renormalized vev $v_{\phi,R}= Z_\phi v_\phi$ stays finite iff $v_\phi$ is finite. A notable exception is the BKT transition, where $Z_\phi$ diverges and renders a finite $v_{\phi,R}$, while $v_{\phi}$ is vanishing.


\subsection{Biconical Fixed Point and Convergence of Local Potential Approximation}

Below $D=4$ dimensions, the model studied in the main text is known to have several non-trivial RG fixed points, most notably, the isotropic, the decoupled, and the biconical fixed point (BFP)~\cite{Fisher:1974,Vicari:2003,Eichhorn:2013mcl}.  
For a specific choice of $N$ only one of these is IR stable and governs the system's multicritical behavior unless additional fine-tuning is performed.
In $D=2+1$, when $N\geq 3$, the decoupled fixed point is found to be stable~\cite{Vicari:2003,Eichhorn:2013mcl}, which was further corroborated by recent conformal bootstrap data~\cite{Kos:2016ysd,Chester:2019ifh,Chester:2020iyt}. The corresponding quantum phase transition in $D\!=\!1\!+\!1$ was recently studied in~\cite{heymans2022quantum}.

We focus on the BFP in $D=2+1$ at $T=0$, which is equivalent to the corresponding statistical models in three spatial dimensions.
Using $\beta$~functions extracted from Eq.~\eqref{eq:RGEeffpotfull} in LPA$({}')n$ for $n\in \{6,8\}$, we calculate fixed-point coordinates and negative eigenvalues $\theta_i$ of the stability matrix, which determine the stability and critical exponents of the BFP. In Fig.~\ref{fig:etasathetas}, we show the anomalous dimensions $\eta_{\phi,\chi}$ and the two leading negative eigenvalues of the stability matrix at the BFP with respect to $N \geq 2$ for $M=1$. Already for $N \geq 5$, the different truncations only differ by $< 1\%$, showing the apparent convergence of the LPA~\cite{Marchais:2013,Eichhorn:2013mcl}. Even for $N=2$, we already obtain good results compared to perturbative 5-loop results, which is summarized in Tab.~\ref{tab:5loopcomp}. For reference, in Tab.~\ref{tab:bfp}, we show the BFP coordinates up to the quartic couplings and the three leading negative eigenvalues of the stability matrix for $N\in \{10,15,20,100\}$.

\begin{table}[t!]
\centering
\begin{tabular}{ |c||c|c|c|c||c| }
\hline
 $N=2$ & LPA${}6$ & LPA${}8$ & LPA${}'6$ & LPA${}'8$ & 5-loop \\ 
 \hline \hline
 $\nu=\theta_1^{-1}$ & 0.675 & 0.605 & 0.679 & 0.627 & 0.70(3) \\
 \hline
 $\omega_1=\theta_2^{-1}$ & 0.766 & 0.762 & 0.727 & 0.707 & 0.79(2) \\
 \hline
 $\eta_\phi$ & 0 & 0 & 0.040 & 0.045 & 0.037(5) \\
 \hline
 $\eta_\chi$ & 0 & 0 & 0.040 & 0.046 & 0.037(5) \\
 \hline
\end{tabular}
\caption{Comparison of two largest eigenvalues of the stability matrix obtained within a LPA with results of a pertubative 5-loop expansion of Ref.~\cite{Vicari:2003}. Note that we set $Z_{\phi, \chi} \equiv 1$ in LPA${}6$ and 8.}
\label{tab:5loopcomp}
\end{table}

\begin{table}[t!]
\centering
\begin{tabular}{ |c||c|c|c|c|c||c|c|c| }
\hline
 $N = 10$ & $\bar{\kappa}_\phi$ & $\bar{\kappa}_\chi$ & $\bar{\lambda}_{\phi}$ & $\bar{\lambda}_{\chi}$ & $\bar{\lambda}_{\phi \chi}$ & $\theta_1$ & $\theta_2$ & $\theta_3$ \\ 
 \hline
 \hline
 LPA${}6$ & 0.25 & 0.10 & 2.62 & 2.54 & -2.34 & 2.02 & 1.06 & 0.61\\
 \hline
 LPA${}8$ & 0.24 & 0.09 & 2.59 & 2.84 & -2.43 & 1.98 & 1.07 & 0.61\\
 \hline
 \hline
 LPA${}'6$ & 0.24 & 0.10 & 2.50 & 2.61 & -2.30 & 1.99 & 1.09 & 0.56\\
 \hline
 LPA${}'8$ & 0.24 & 0.09 & 2.47 & 2.81 & -2.34 & 1.95 & 1.09 & 0.57\\
 \hline
 \hline
 \hline
 $N = 15$ & $\bar{\kappa}_\phi$ & $\bar{\kappa}_\chi$ & $\bar{\lambda}_{\phi}$ & $\bar{\lambda}_{\chi}$ & $\bar{\lambda}_{\phi \chi}$ & $\theta_1$ & $\theta_2$ & $\theta_3$ \\ 
 \hline
 \hline
 LPA${}6$ & 0.35 & 0.12 & 1.81 & 1.73 & -1.64 & 2.02 & 1.04 & 0.71\\
 \hline
 LPA${}8$ & 0.35 & 0.11 & 1.81 & 1.94 & -1.72 & 1.99 & 1.05 & 0.72\\
 \hline
 \hline
 LPA${}'6$ & 0.35 & 0.12 & 1.76 & 1.79 & -1.64 & 2.01 & 1.06 & 0.67\\
 \hline
 LPA${}'8$ & 0.34 & 0.11 & 1.75 & 1.95 & -1.69 & 1.97 & 1.06 & 0.69\\
 \hline
 \end{tabular}
 \begin{tabular}{ |c||c|c|c|c|c||c|c|c| }
 \hline
 $N = 20$ & $\bar{\kappa}_\phi$ & $\bar{\kappa}_\chi$ & $\bar{\lambda}_{\phi}$ & $\bar{\lambda}_{\chi}$ & $\bar{\lambda}_{\phi \chi}$ & $\theta_1$ & $\theta_2$ & $\theta_3$ \\ 
 \hline
 \hline
 LPA${}6$ & 0.45 & 0.14 & 1.39 & 1.32 & -1.27 & 2.02 & 1.03 & 0.77\\
 \hline
 LPA${}8$ & 0.45 & 0.13 & 1.38 & 1.48 & -1.33 & 1.99 & 1.04 & 0.78\\
 \hline
 \hline
 LPA${}'6$ & 0.45 & 0.13 & 1.36 & 1.37 & -1.27 & 2.01 & 1.04 & 0.74\\
 \hline
 LPA${}'8$ & 0.44 & 0.13 & 1.35 & 1.49 & -1.31 & 1.97 & 1.05 & 0.75\\
 \hline
 \hline
 \hline
 $N=100$ & $\bar{\kappa}_\phi$ & $\bar{\kappa}_\chi$ & $\bar{\lambda}_{\phi}$ & $\bar{\lambda}_{\chi}$ & $\bar{\lambda}_{\phi \chi}$ & $\theta_1$ & $\theta_2$ & $\theta_3$ \\ 
 \hline
 \hline
 LPA${}6$ & 1.94 & 0.28 & 0.29 & 0.29 & -0.28 & 2.00 & 1.01 & 0.94\\
 \hline
 LPA${}8$ & 1.92 & 0.26 & 0.29 & 0.31 & -0.29 & 1.99 & 1.01 & 0.95\\
 \hline
 \hline
 LPA${}'6$ & 1.92 & 0.35 & 0.29 & 0.16 & -0.21 & 2.00 & 1.01 & 0.95\\
 \hline
 LPA${}'8$ & 1.91 & 0.32 & 0.29 & 0.17 & -0.21 & 1.99 & 1.01 & 0.96\\
 \hline
\end{tabular}
\caption{BFP coordinates and leading negative eigenvalues of the stability matrix for $N=10,15,20,100$ and $M = 1$.}
\label{tab:bfp}
\end{table}

We note that for statistical models in two spatial dimensions the calculation of critical exponents is more challenging as all interactions $\sim \Phi^{2n}$ are canonically relevant and should be accounted for. 
This can be done based on the flow of the full potential~\cite{PhysRevLett.110.141601,PhysRevE.101.042113,Jentsch:2022,Dupuis:2021}.
Yet, within our approach, we find that finite-order LPA$(')$ already describes the Ising transition and CHMW faithfully, which we will use to proceed to $T>0$ in $D=2+1$.


\section{Comment on discontinuity line in phase diagram and numerical stability}

\begin{figure}[h!]
    \centering
    \includegraphics[scale=0.87]{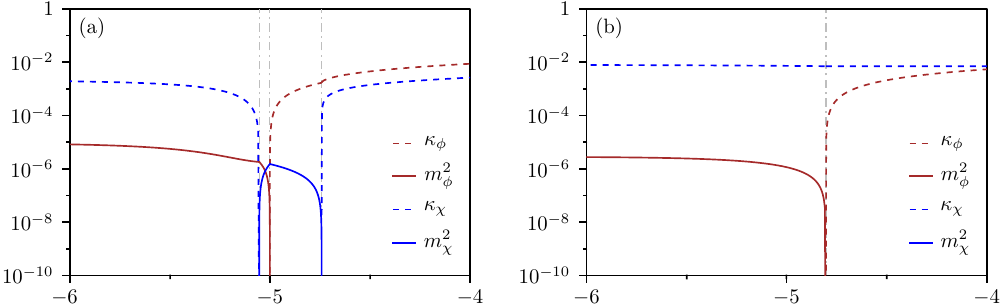}
    \caption{Exemplary flows left (a) and right (b) of the discontinuity line appearing in the phase diagram shown in Fig.~3 of the main text.}
    \label{fig:flowdisc}
\end{figure}

Upon close inspection, Fig.~3 of the main text reveals a faint transition line in the SSB phase emerging from the QCP, indicating a discontinuous jump of the (squared) vacuum expectation value $\kappa_\chi^\mathrm{IR}$.
In Fig.~\ref{fig:flowdisc}, we show a flow close to the left~(a) and close to the right~(b) of the line. 
The appearing line seems to separate two classes of flows, namely, flows where $\mathbb{Z}_2$ symmetry is restored on intermediate RG scales, cf. Fig.~\ref{fig:flowdisc}(a), and flows where $\mathbb{Z}_2$ is broken on all scales, cf. Fig.~\ref{fig:flowdisc}(b). The two phases have the same IR behaviour, suggesting that the jump is due to an accumulated numerical error from the switching of different regimes of the potential. 
This could be resolved by more advanced methods in the future~\cite{Borchardt:2015ps, Borchardt:2015ps2,Borchardt:2016,Grossi:2019urj}.

We also note that for certain choices of small $N$ and LPA$(')n$ there are numerical instabilities occurring in some regions of the phase diagram while integrating the RG equations. 
We can attribute this to be an artefact of the finite-order of the LPA.
In fact, for large $N$, where higher-dimensional operators can be expected to be suppressed, no numerical instabilities occur within LPA${}(')6/8$ and the solution of the FRG flow is completely stable.
For smaller $N = \mathcal{O}(10)$, we were only able to integrate to the IR within LPA${}(')6$, as the local expansion seems to break down during RG flow in LPA${}(')8$. 
Similar behavior has already been observed in previous works, see, e.g., Refs.~\cite{Scherer:2013many, Torres:2020}.  
A possibility of by-passing those issues is to resolve the full potential on a grid in field space by employing more advanced numerical methods~\cite{Borchardt:2015ps, Borchardt:2015ps2,Borchardt:2016,Grossi:2019urj}. This is, however, beyond the scope of the present work.

\end{document}